\documentclass[twocolumn,secnumarabic,amsmath,amssymb,nofootinbib,floatfix]{revtex4-1}

\setcitestyle{super}

\usepackage{graphics}      
\usepackage{graphicx}     
\usepackage{epstopdf}
\usepackage{url}           
\usepackage{bm}            
\usepackage[caption=false]{subfig}    
\usepackage{float}
\usepackage{amsmath}
\usepackage[usenames,dvipsnames]{xcolor}

\usepackage{epstopdf}

\newcommand{\REV}[1]{{\color{black} #1}}

\begin{document}
\title{Self-propelled Worm-like Filaments: Spontaneous Spiral Formation, Structure, and Dynamics}
\author{Rolf E. Isele-Holder}
\email{r.isele-holder@fz-juelich.de}
\affiliation{Theoretical Soft Matter and Biophysics,
Institute of Complex Systems
and Institute for Advanced Simulations,
Forschungszentrum J\"ulich GmbH, 52425 J\"ulich, Germany}

\author{Jens Elgeti}
\email{j.elgeti@fz-juelich.de}
\affiliation{Theoretical Soft Matter and Biophysics,
Institute of Complex Systems
and Institute for Advanced Simulations,
Forschungszentrum J\"ulich GmbH, 52425 J\"ulich, Germany}

\author{Gerhard Gompper}
\email{g.gompper@fz-juelich.de}
\affiliation{Theoretical Soft Matter and Biophysics,
Institute of Complex Systems
and Institute for Advanced Simulations,
Forschungszentrum J\"ulich GmbH, 52425 J\"ulich, Germany}

\date{\today}

\begin{abstract}
Worm-like filaments that are propelled homogeneously along their tangent
vector are studied by Brownian dynamics simulations.
Systems in two dimensions are investigated, corresponding to
filaments adsorbed to interfaces or surfaces.
A large parameter space covering weak and strong
propulsion, as well as flexible and stiff filaments
is explored.
For strongly propelled and flexible filaments, the
free-swimming filaments spontaneously form stable spirals. 
The propulsion force has a strong impact on dynamic properties,
such as the rotational and translational mean square displacement
and the rate of conformational sampling.
In particular, when the active self-propulsion dominates thermal diffusion,
but is too weak for spiral formation, the rotational diffusion coefficient has
an activity-induced contribution given by $v_c/\xi_P$, where $v_c$ is the contour
velocity and $\xi_P$ the persistence length.
In contrast, structural properties are hardly affected
by the activity of the system, as long as no spirals form.
The model mimics common features of
biological systems, such as microtubules and
actin filaments on motility assays or slender
bacteria, and artificially designed microswimmers.
\end{abstract}

\maketitle

\section{Introduction}
\label{s:introduction}

Its importance in biology and its enormous potential impact
in technical applications makes active soft matter a
field of rapidly growing interest and progress.\cite{Elgeti.2015,
Marchetti.2013, Cates.2011}
Flexible slender bodies are of particular
importance.
The majority of natural swimmers propel themselves using flexible, hair-like
structures like cilia and flagella.\cite{Elgeti.2015}
Another important example are actin filaments and microtubules,
major constituents of the cytoskeleton, whose capability to
buckle decisively controls the mechanical properties
of the cell body.\cite{Rodriguez.2003}
Flexibility is the crucial ingredient for the formation
of small-scale spirals\cite{RashedulKabir.2012} and
possibly also for large-scale swirls\cite{Sumino.2012}
of microtubules on motility assays.
Even the structure of slender bacteria can be dominated by their
flexibility.\cite{Lin.2014}
Elextrohydrodynamic convection can propel colloid chains
because they are flexible,\cite{Sasaki.2014} just as
the swimming mechanism of assembled magnetic beads in
an oscillating external magnetic field is possible because
of the swimmer's flexibility.\cite{Vach.2015}
Flexibility is of course also the feature that allows for
the instabilities leading to cilia-like
beating in artificially bundled microtubules.\cite{Sanchez.2011,
Sanchez.2012}

Despite its importance, the number of theoretical studies
of active agents that incorporate flexibility is still relatively small, and can roughly be
subdivided into works that focus on buckling phenomena
and on free-swimming agents.
Symmetry breaking instabilities leading to
rotation and beating motion of active filaments
on motility assays can be described with a phenomenological ordinary differential equation for the filaments.\cite{Sekimoto.1995}
The propulsion force of motor proteins has been predicted based on
a Langevin model for buckled, rotating actin filaments and microtubules.\cite{Bourdieu.1995}
Numerical studies with Lattice-Boltzmann simulations and Brownian or multi-particle
collision dynamics have demonstrated that clamped of pinned filaments composed of
stresslets or propelled beads can show cilia-like beating or rotation.\cite{Laskar.2013, Chelakkot.2014}

The behaviour of free-swimming actin filaments on motility assays
was reproduced in early numerical studies using the Langevin
equation.\cite{Farkas.2002}
However, it was only recently
that theoretical study of flexible, active filaments that can move freely
has received significant attention.
Lattice-Boltzmann simulations reveal that 
spontaneous symmetry breaking in chains of stresslets
can lead to rotational or translational filament motion.\cite{Jayaraman.2012}
Brownian dynamics simulations of short self-propelled filaments
suggest that different types of motion occur for single filaments\cite{Jiang.2014_1}
and that spontaneous rotational motion can arise for pairs of filaments.\cite{Jiang.2014_2}
A combination of Brownian dynamics simulations and analytic
theory shows that shot noise in worm-like filaments
leads to temporal superdiffusive filament movement
and faster-decaying tangent-tangent correlation functions.\cite{Ghosh.2014}
Finally, chains of active colloids connected by springs have the same
Flory exponent but a different prefactor of the scaling law
compared to chains of passive colloids, as shown
recently both analytically for beads
without volume exclusion and numerically with Brownian dynamics simulations
for beads with volume exclusion.\cite{Kaiser.2015}

The free-swimming behaviour of a worm-like filament that is tangentially
propelled with a homogeneous force
is still unexplored and is the subject of this work.
The model is introduced in Section~\ref{s:methods}.
Results for the structural and dynamic properties over a wider range
or propulsion forces and filament flexibilities are presented in Section~\ref{s:results}.
We find that the filament can spontaneously form spirals, which
is the mechanism that dominates the behaviour for large propulsion forces.
The relevance of our observations for natural and
artificial active agents is discussed in Section~\ref{s:discussion}.
We present our conclusions in Section~\ref{s:conclusions}.

\section{Model and Methods}
\label{s:methods}

We study a single, active, worm-like filament,
which is modelled as a sequence of $N+1$~beads
connected via stiff springs. The overdamped equation of motion
is given by
\begin{equation}
  \gamma \dot{\mathbf{r}_i} = -\nabla_i U + \mathbf F_\mathrm{k_BT}^{(i)} + \mathbf F_p^{(i)},
\end{equation}
where $\mathbf{r}_i$ are the coordinates of bead $i$,
$\gamma$ is the friction coefficient, $U$ is the configurational energy,
$\mathbf{F}_{k_BT}^{(i)}$ is the thermal noise force, and $\mathbf{F}_p^{(i)}$
is the active force that drives the system out of equilibrium.
The configurational potential energy
\begin{equation}
  U =  U_\mathrm{bond} + U_\mathrm{angle} + U_\mathrm{EV}
\end{equation}
is composed of a bond contribution between neighbouring beads
\begin{equation}
  U_\mathrm{bond} = \frac{k_S}{2} \sum_{i=1}^N (|\mathbf{r}_{i,i+1}| - r_0)^2,
\end{equation}
a bending energy
\begin{equation}
  U_\mathrm{angle} = \frac{\kappa}{4} \sum_{i=1}^{N-1} (\mathbf{r}_{i,i+1} - \mathbf{r}_{i+1,i+2})^2,
\end{equation}
and an excluded volume term modelled with repulsive Lennard-Jones interactions
\begin{eqnarray}
  U_\mathrm{EV} & = & \sum_{i=1}^N\sum_{j > i}^{N+1} u_\mathrm{EV}(r_{i,j}),  \\
  u_\mathrm{EV}(r) & = & 
    \left\{
      \begin{array}{lr}
        4\epsilon \left[ \left(\frac{\sigma}{r} \right)^{12}
        - \left(\frac{\sigma}{r} \right)^{6} \right] + \epsilon, & r < 2^{1/6}\sigma \\
        0, &  r \geq 2^{1/6} \sigma,
      \end{array}
    \right.
\end{eqnarray}
where $\mathbf{r}_{i,j} = \mathbf{r}_i - \mathbf{r}_j$
is the vector between
the position of the beads $i$ and $j$, $k_S$ is the spring
constant for the bond potential, $r_0$ is the equilibrium
bond length, $\kappa$ is the bending rigidity,
and $\epsilon$ and $\sigma$ are the characteristic volume-exclusion
energy and effective
filament diameter (bead size).

The drag force $\gamma \dot{\mathbf{r}_i}$ is the velocity of
each bead times the friction coefficient $\gamma$. The thermal force
$\mathbf{F}_{k_BT}^{(i)}$ is modelled as white noise with zero mean
and variance $2k_BT \gamma / \Delta t$ as described in
Ref.~\citenum{Dunweg.1991}.
\REV{Note that hydrodynamic interactions (HI) are not
included in our model. The model is thus in particular valid for
(i) neutral swimmers, for which HI are known to be of minor importance,
\cite{Downton.2009, Goetze.2010, Zoettl.2014}
(ii) swimmers near a wall, where HI is of less
importance,\cite{Elgeti.2009, Drescher.2011}
and (iii) microorganisms that glide on a surface, such as nematodes like
\it{C. elegans}.\cite{Gray.1964, Korta.2007}}

\begin{figure}[t]
  \centering
  \includegraphics[width=2in]{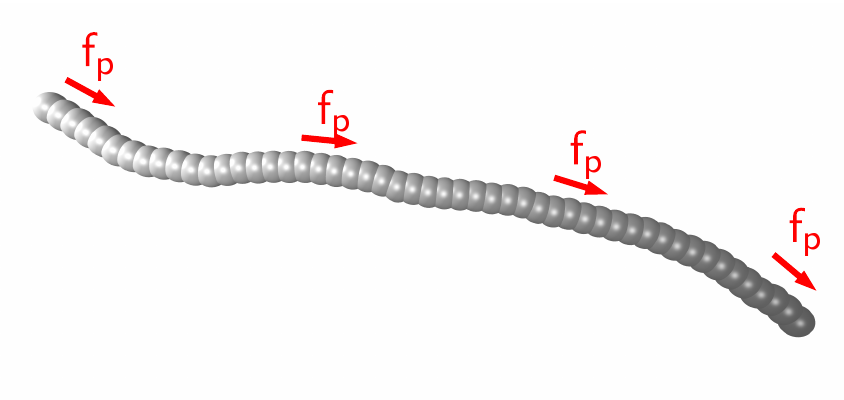}
  \caption{Filament model: Beads are connected via stiff springs. The
    active force acts tangentially along all bonds. Colour gradient
    indicates the force direction.}
  \label{f:polymer_model}
\end{figure}

Without propulsion force, $\mathbf{F}_p^{(i)} = 0$, the model
matches the well-known worm-like
chain model for semi-flexible
polymers.\cite{Kratky.1949, Saito.1967} For active filaments, we use a force
per unit length $f_p$ that acts tangentially along all bonds, i.e.,
\begin{equation}
  \mathbf{F}_p = \sum_i^N f_p\mathbf{r}_{i,i+1},
\end{equation}
as illustrated in Fig.~\ref{f:polymer_model}.
The force along each bond is distributed equally onto both adjacent beads.

We consider systems with parameters chosen such that (i) $k_S$ is sufficiently
large that the bond length is
approximately constant $r_0$, that (ii) the local filament
curvature is low such that the bead discretization does not violate the
worm-like polymer description, and that (iii) the thickness of the
chain has negligible impact on the results. When these requirements
are met, the system is fully characterized by two dimensionless numbers,
\begin{eqnarray}
  \xi_P/L & = & \frac{\kappa}{k_BTL},           \\
  Pe & = & \frac{v_cL}{D_t} = \frac{f_pL^2}{k_BT},
\end{eqnarray}
where $L=Nr_0$ and $\xi_P$ are the length and persistence length of the
chain, respectively. $\xi_P/L$ is a measure for the bending rigidity of
the filament. The P\'eclet number $Pe$ is the ratio of convective
to diffusive transport and measures the degree of activity. For
its definition, we use that the filament has a contour velocity
$v_c = f_p  / \gamma_l $,
and that the translational diffusion coefficient $D_t = k_BT/\gamma_l L$,
where we have introduced the friction per unit length $\gamma_l = \gamma (N+1) /L$.

The ratio of these numbers
\begin{equation}
  \mathfrak{F} = PeL/\xi_P = \frac{f_pL^3}{\kappa},
\end{equation}
which we call the flexure number, provides a ratio of activity
to bending rigidity. Previous studies showed that this number
is decisive for buckling instabilities of active
filaments.\cite{Sekimoto.1995,Chelakkot.2014} It will be shown
below that this is also a determining quantity for spiral stability
and rotational diffusion.

\begin{figure*}[t]
  \centering
\subfloat[][polymer regime]{\includegraphics[width=5.7cm]{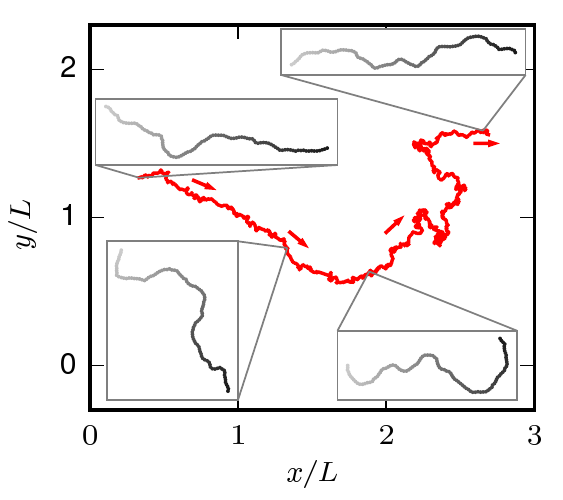}}
\subfloat[][weak spiral regime]{\includegraphics[width=5.7cm]{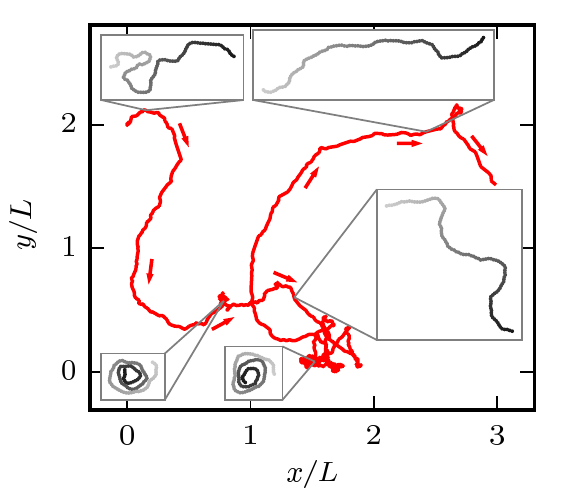}}
\subfloat[][strong spiral regime]{\includegraphics[width=5.7cm]{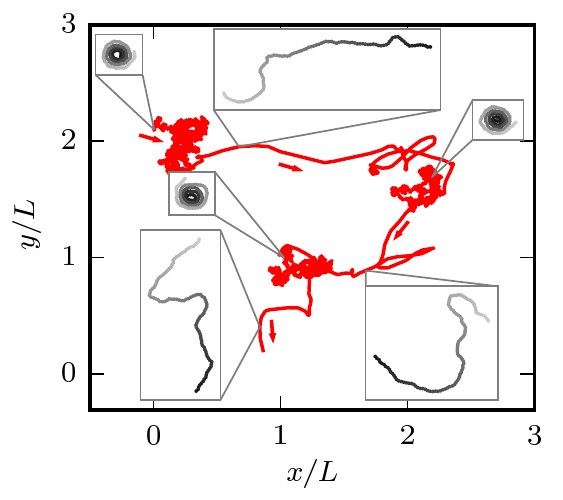}}

  \caption{Trajectories of the center of mass of the filament (red) and
    filament configurations from selected snapshots (grayscale, leading tip black). Arrows point
    in the direction of movement. $\xi_P/L=0.2$ \REV{(with $N=100$)}
    in all plots; $Pe$ increases from left to right.
    At $Pe=200$ (left), there is not sign of spiral formation.
    At $Pe = 1000$ (middle) spirals form occasionally, but the overall behaviour
    is dominated by an elongated chain. 
    At $Pe = 5000$ (right), the spiral state is predominant.
    The chain has a directed motion in the elongated state. In the spiral state,
    the translational motion is almost purely diffusive. This leads to separated clusters in the trajectories
    for simulations in the strong spiral regime, visible for example
    in the upper left of the right image.
    The length of the depicted trajectories corresponds to
    approximately $0.13 \tau$ (left), $0.13 \tau$ (middle), and $0.6 \tau$ (right).}
  \label{f:trajectories}
\end{figure*}

Simulations were performed in two dimensions, where volume exclusion interactions
have major importance. Equations of motions were integrated using an
Euler scheme.
Simulation parameters and results are reported in dimensionless form, 
where length are measured in units of the filament length $L$, 
energies in units of the thermal energy $k_BT$, and time in units
of the characteristic time for the filament to diffuse its own
body length
\begin{equation}
  \tau = L^3 \gamma_l /4k_BT.
\end{equation}
In our simulations we used $k_S = 4000$\,$k_BT/r_0^2$,
$r_0=\sigma=L/N$, and $\epsilon=k_BT$
\REV{if not stated otherwise.} A large parameter space for $Pe$ and $\xi_P/L$ was explored
by varying $f_p$, $N$, and $\kappa$. \REV{$N$ was varied in the range
from 25 to 200 from the highest to the lowest $\xi_P/L$.
Almost all simulations were run for more than $5\,\tau$.
An initial period of the simulation output is discarded
in the analysis.}
The timestep $\Delta t$ was adjusted
to the remaining settings to ensure stable simulations.
Unless explicitly mentioned, results refer to simulations
that were started with a perfectly straight conformation.

All simulations were performed using the LAMMPS molecular simulation
package\cite{Plimpton.1995} with in-house modifications to describe
the angle potential, the propulsion forces, and to solve the
overdamped equations of motion.

\section{Results}
\label{s:results}

\begin{figure}[t]
  \centering

\subfloat[Spontaneous spiral
formation]{\includegraphics[width=8.3cm]{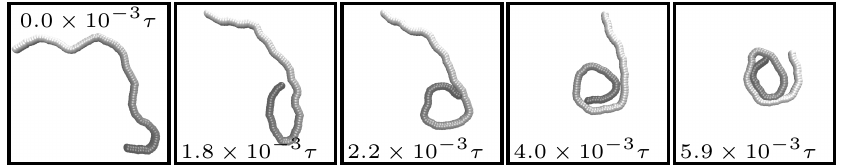}}

\subfloat[Spontaneous spiral
break-up]{\includegraphics[width=8.3cm]{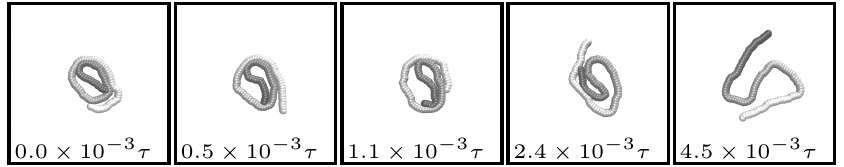}}

\subfloat[Spiral break-up by widening]{\includegraphics[width=8.3cm]{figs/uncoiling.pdf}}

  \caption{Spiral formation and break-up mechanisms.
    Numbers in each panel provide the
    elapsed time. Leading tip is black. \REV{$N=100$ in each case.
    (a,b) $\xi_P/L = 0.4$, $Pe = 10\,000$, $\mathfrak{F} = 25\,000$. (c) $\xi_P/L = 1.0$, $Pe = 900$, $\mathfrak{F} = 900$.}
    }
  \label{f:coil_uncoil}
\end{figure}

\begin{figure*}[t]
  \centering
  \begin{minipage}[t]{5.7cm}
    \includegraphics[width=5.7cm]{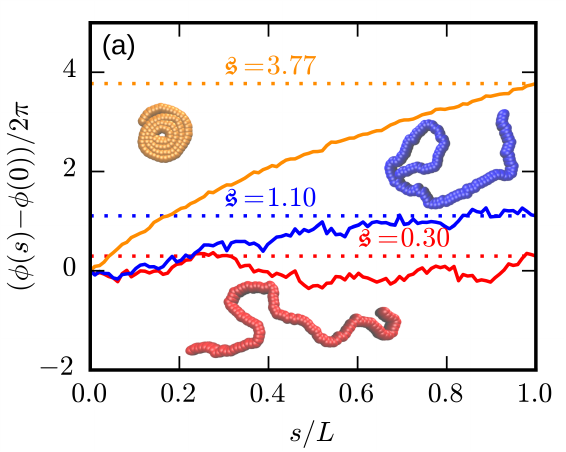}
    \includegraphics[width=5.7cm]{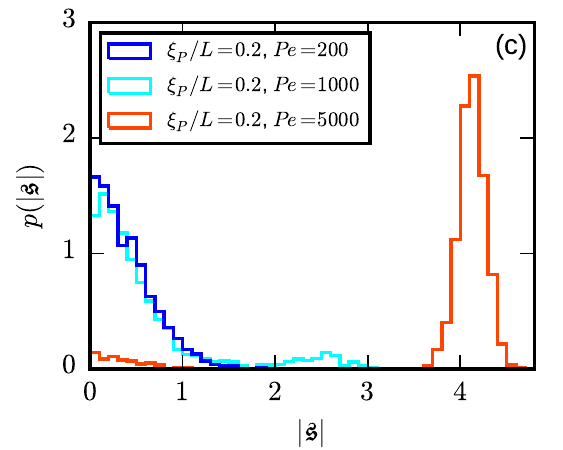}
  \end{minipage}
  \begin{minipage}[t]{5.7cm}
    \includegraphics[width=5.7cm]{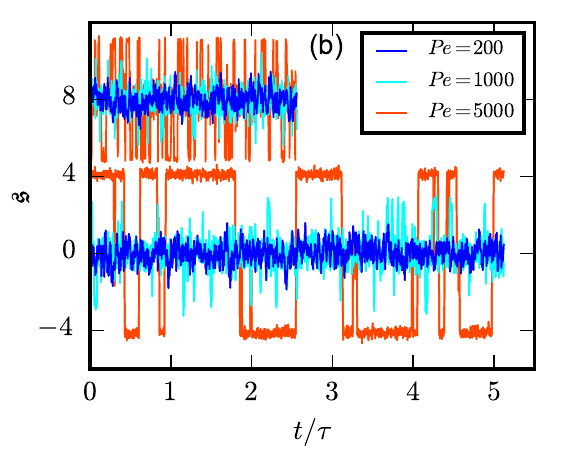}
    \includegraphics[width=5.7cm]{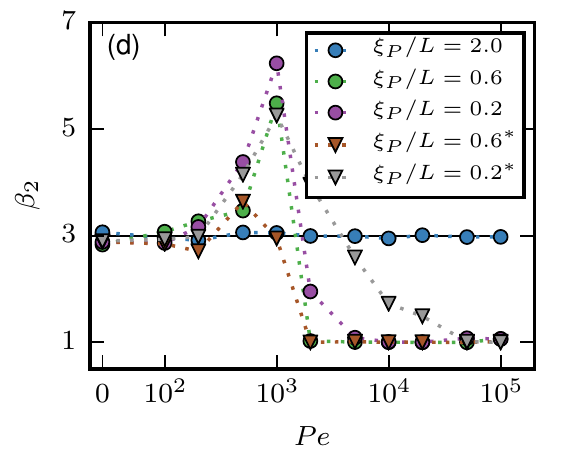}
  \end{minipage}
  \begin{minipage}[c]{5.7cm}
    \vspace{0.5cm}
    \includegraphics[width=5.7cm]{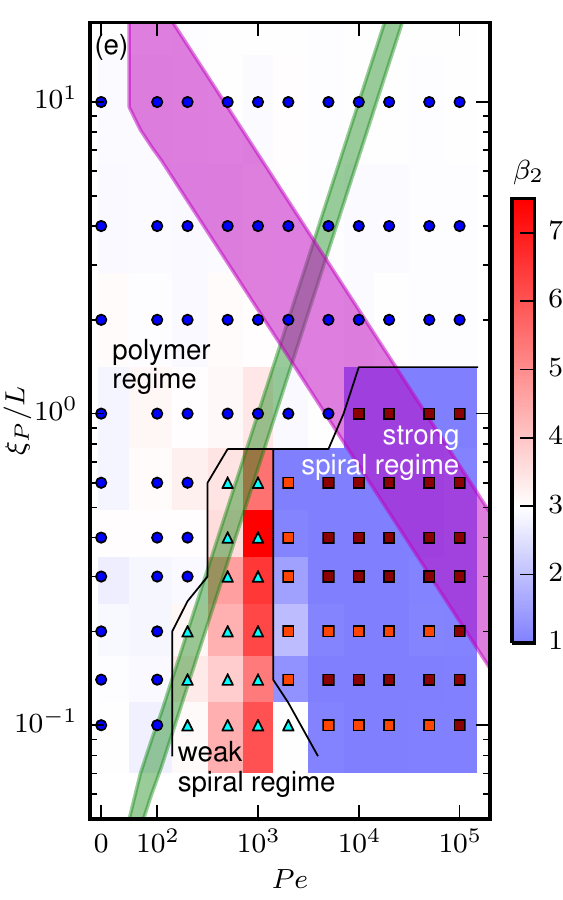}
  \end{minipage}
  \caption{
    (a) Definition of the spiral number $\mathfrak{s}$. The measure can
    effectively distinguish between elongated (red), weakly wound-up (blue),
    and strongly wound-up (orange) conformations.
    (b) Evolution of the spiral number $\mathfrak{s}$ for the examples given in
    Fig.~\ref{f:trajectories} \REV{with $\xi_P/L = 0.2$ (N = 100)}. \REV{Lines centered around $\mathfrak{s} =
    0$ display results for $\sigma = L/N$; lines centered around $\mathfrak{s} =
    8$ (shifted upwards for clarity) show results for $\sigma
    = 2L/N$.}
    (c) Probability distributions of the absolute value of the spiral number $p(|\mathfrak{s}|)$ for the same
    examples with $\sigma = L/N$.
    (d) Kurtosis $\beta_2$ for selected filament rigidities. \REV{Circles:
    $\sigma=L/N$, triangles: $\sigma = 2L/N$.}
    (e) Phase diagram. Background colour: kurtosis. Blue circles:
    polymer regime. Cyan triangles: weak spiral regime. Light
    and dark red squares: strong spiral regime. For the dark red squares, spirals did not break up during the
    simulations once formed. Black lines are a guide to the eye. Green area: threshold for spiral stability
    against break-up by widening. Spirals above this threshold will unfold by widening, spirals
    below will not. \REV{Purple}: parameter space that can be obtained by actin filaments on a myosin carpet at $T=300$\,K using parameters
    for $f_p$ and $\kappa$ from Ref.~\citenum{Sekimoto.1995}.
  }
  \label{f:coil_phases}
\end{figure*}

The characteristic filament behaviour depends on its bending rigidity
and activity and can be divided into three regimes
(see Fig.~\ref{f:trajectories}).
At low $Pe$ or high $\xi_P/L$, the ``polymer regime'', the 
active filament structurally resembles the passive filament with
$Pe = 0$. The main difference compared to the passive filament is that the active force drives the
filament along its contour,
leading to a directed translational
motion --- we name this characteristic movement ``railway motion''.
At high $Pe$ and low $\xi_P/L$, the
filament spontaneously winds up to a spiral.
The ``spiral state'' is characterized by
ballistic rotation but only diffusive translation.
Spirals can spontaneously break up. Their lifetime
determines whether spiral formation has a minor
impact on the overall filament behaviour,
the ``weak spiral regime'' at intermediate $Pe$, or whether
spirals are dominating, the ``strong spiral regime'' at large $Pe$.
Because spiral formation has a major impact on both the structure and
the dynamics, features related to spiral formation are addressed first.
Structural and dynamic properties of the elongated and spiral state
are presented afterwards.

\subsection{Spiral Formation}
\label{s:coil_formation}

The processes that lead to the formation and break-up of spirals
are depicted in Fig.~\ref{f:coil_uncoil}.
Spontaneous spiral formation
(cf. Fig.~\ref{f:coil_uncoil}a) results from the leading tip 
colliding with a subsequent part of the chain.
Volume exclusion then forces the tip
to bent. By further forward movement, the chain winds to a spiral.
Two spiral break-up mechanisms occurred in our simulations.
The first is the thermally activated mechanism in
Fig.~\ref{f:coil_uncoil}b. The leading tip of the wound-up
chain spontaneously changes direction
and the spiral deforms. This break-up mechanism requires
strong local bending and is therefore 
predominant for small $\xi_P$.
The second mechanism is spiral break-up by widening and is
depicted in Fig.~\ref{f:coil_uncoil}c. The bending potential
widens the spiral until the leading tip looses contact to the filament end.
This break-up mechanism is predominant when $\xi_P$ is
too large for spontaneous spiral break-up. Because high
stiffness is also unfavourable for spiral formation, spiral
break-up by widening was almost exclusively observed in simulations
that started with a spiral configuration.

To understand spiral formation more quantitatively, we introduce the spiral number
\begin{equation}
  \mathfrak{s} = (\phi(L) - \phi(0))/2\pi,
\end{equation}
where $\phi(s)$ is the bond orientation at position $s$ along
the contour of the filament, as measure for
the instantaneous chain configuration. The definition is
illustrated for three sample structures in
Fig.~\ref{f:coil_phases}a.
It effectively measures how
often the filament wraps around itself. The time evolution of
$\mathfrak{s}$ is depicted in Fig.~\ref{f:coil_phases}b for the same $Pe$ and
$\xi_P/L$ as in Fig.~\ref{f:trajectories}. At $Pe = 200$, $\mathfrak{s}$ is
always close to zero. At $Pe = 1000$, $\mathfrak{s}$ behaves similarly, except that
peaks with larger values for $|\mathfrak{s}|$ occur occasionally, i.e., when spirals
with a short lifetime form. At $Pe = 5000$, extended plateaus develop
at large $|\mathfrak{s}|$. The spirals are wound up stronger and have a much longer
lifetime. 

Probability distributions $p(|\mathfrak{s}|)$ are depicted in
Fig.~\ref{f:coil_phases}c. For the simulation without
spirals ($Pe=200$), the histogram resembles the right half of a
Gaussian distribution. For the simulation in the weak spiral regime,
$p(\mathfrak{s})$ is similar for low $|\mathfrak{s}|$, but also has a small peak
at $|\mathfrak{s}| \approx 2 - 3$. For strong spiral formation at $Pe = 5000$, 
$p(|\mathfrak{s}|)$ has only a small peak at low $|\mathfrak{s}|$, which corresponds
to the elongated state, and a large peak at large $|\mathfrak{s}|$, the
predominating spiral state.
It turns out that the different regimes can be well distinguished
by the kurtosis
\begin{equation}
  \beta_2 = \left\langle \left( \frac{\mathfrak{s} - \left\langle \mathfrak{s} \right\rangle}{\sigma_\mathfrak{s}} \right)^4 \right\rangle,
  \label{e:kurtosis}
\end{equation}
where $\langle \dots \rangle$ denotes the ensemble average
and $\sigma_\mathfrak{s}$ is the standard deviation of $\mathfrak{s}$.
Results for the kurtosis are shown in
Fig.~\ref{f:coil_phases}d for selected
$\xi_P/L$. $\beta_2 \approx 3$ in the polymer regime, as
expected for Gaussian distributions. In the weak spiral regime, the 
small peak at larger values increases the numerator in Eq.~(\ref{e:kurtosis})
and has only a weak impact on $\sigma_\mathfrak{s}$, leading to an increase of
$\beta_2$. When the spiral state is dominating, $\sigma_\mathfrak{s}$ grows drastically,
resulting in a much smaller kurtosis $\beta_2$.
Note that to reduce statistical uncertainties we symmetrized
the $\mathfrak{s}$-distribution in the computation of $\beta_2$ by counting
each measured $|\mathfrak{s}|$ as $+\mathfrak{s}$ and $-\mathfrak{s}$
\REV{and only used data from the spiral state for simulations that do not
show spiral break-up.}

With the kurtosis as measure to characterize spiral formation,
a phase diagram
can be constructed as depicted in
Fig.~\ref{f:coil_phases}e. Low filament rigidity $\xi_P$ and
high propulsion $Pe$ is beneficial for spiral formation. In particular,
for a fixed propulsion strength per unit length, any chain will form
spirals if it is sufficiently long, because increasing the chain
length without modifying any other parameter corresponds to moving to
the lower right in the phase diagram.

\REV{The dimensionless numbers $\xi_P/L$ and $Pe$ completely characterize the system if
  the filament diameter --- or the filament aspect ratio --- is of
  minor importance. This is the true in the entire polymer regime,
  where volume-exclusions interactions hardly come into play because
  of the elongated chain structure. For the spiral regimes, the aspect
  ratio has an impact on the structure of the spiral and does in this
  way influence the results. Which features of the spiral regime can
  be approximated well by the dimensionless numbers can be understood
  from the spiral formation and break-up mechanisms.
  The aspect ratio is hardly relevant for} spiral formation and spiral
  break-up by widening, where the decisive moments are when the
  filament tip collides with subsequent parts of the chain, or when it
  looses contact to the chain end, respectively. That break-up by
  widening is characterized well by the dimensionless numbers is also
  confirmed by a series of simulations that we start from a spiral
  configuration in which we vary $N$, $f_p$, $\kappa$, and $k_BT$. It
  turns out that spirals will break up by widening if
\begin{equation}
  \mathfrak{F} \lesssim 1000 - 1500.
  \label{e:widening}
\end{equation}

In contrast, spontaneous spiral break-up by a change of orientation of the
leading tip is dependent on a strong local curvature close
to the tip and the structure of the spiral, which in turn
is dependent on the filament diameter. The dimensionless
description does therefore
not provide a full characterization of the strong spiral
regime, where spontaneous spiral
break-up is the only mechanism to escape the spiral state.
This is also confirmed by results for
spirals that never broke up (cf. dark red squares in
Fig.~\ref{f:coil_phases}e), results for $\xi_P/L=0.2$
and $\xi_P/L=0.14$ show non-monotonic behaviour in the direction
of $\xi_P/L$.
This is a result of a combination of that the 
dimensionless description is only partially valid in this regime
and that we chose $N=200$ for $\xi_P/L < 0.2$ but
$N=100$ for $0.2 \geq \xi_P/L \geq 2.0$ 
in our simulations, \REV{i.e., the aspect ratio $L/\sigma$ is halved
in our simulations for filaments with $\xi_P/L < 0.2$.}

\REV{Finally, the bead discretization with the chosen parameters
  favours a staggered arrangement of beads of contacting parts of
  the filament,\cite{Yang.2010, Abkenar.2013} which implies an
  effective sliding friction between these parts. To study the
  importance of this effect, we increase the
  diameter of the beads at fixed bond length so that neighboring beads
  are heavily overlapping, which leads to a strongly smoothened
  interaction potential. Results for an increased
  diameter $\sigma = 2L/N$ are shown
  in Fig.~\ref{f:coil_phases}b~and~d. We find that the spiral formation
  frequency is hardly affected by smoothening the filament surface. In
  contrast, spontaneous spiral break-up is largely alleviated for the
  smoother filament, leading to decreased spiral life-time, as can be
  seen from the evolution of $\mathfrak{s}$ for $Pe=5000$ in
  Fig.~\ref{f:coil_phases}b. 
  Smoother filaments thus show a qualitatively similar phase behaviour
  with slightly moved phase boundaries.}

\subsection{Structural Properties}
\label{s:statics}

\begin{figure}[t]
  \centering
  \includegraphics[width=8.3cm]{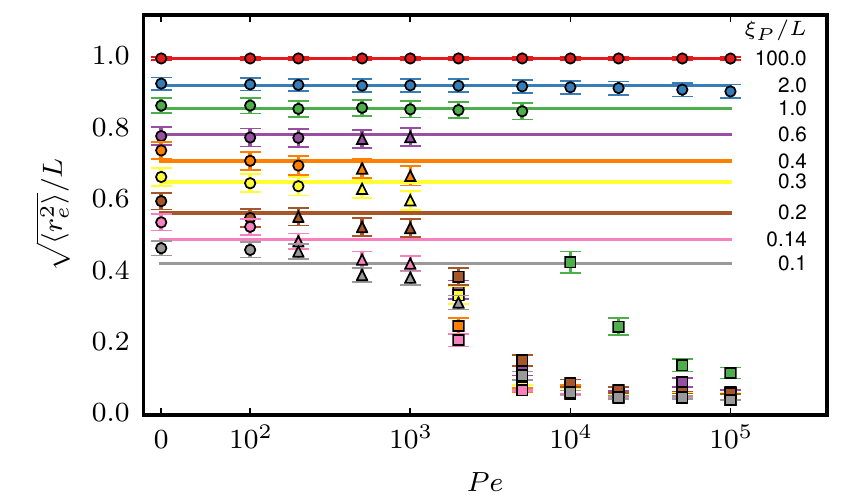}
  \caption{Symbols: Mean end-to-end distance
  $\sqrt{\langle r_{e}^2   \rangle}$ over $Pe$ for different
  values of $\xi_P/L$.
  Solid lines: $\sqrt{\langle r_{e}^2   \rangle}$ 
  as predicted from the Kratky-Porod model. The symbol shape
  indicates the region in the phase diagram:
  circle: polymer regime; triangle: weak spiral regime, squares: strong spiral regime. $N$ varies from 25 to 200 from large to small $\xi_P/L$.
  } 
  \label{f:re2e}
\end{figure}

The structural properties of the filament conformations
can be best understood
from the end-to-end vector $r_{e}$, as depicted in
Fig.~\ref{f:re2e}. As long as no spirals form, simulation
results are in good agreement with the Kratky-Porod model
(valid for
worm-like, non-active polymers without volume-exclusion interactions)
that predicts\cite{Kratky.1949, Saito.1967}
\begin{equation}
  \frac{\langle r_{e}^2 \rangle}{L^2} = 2\frac{\xi_P}{L} - 2\left(\frac{\xi_P}{L}\right)^2 \left(1 - e^{-L/\xi_P} \right)
\end{equation}
in two dimensions.
At low $\xi_P/L$, volume-exclusion interactions lead to slight
deviations between the Kratky-Porod model and the simulation results.
Strong deviations between the Kratky-Porod model and simulation
results only occur in the strong spiral region in the phase
diagram. The same trend was observed for the tangent-tangent
correlation function, the radius of gyration, and the static 
structure factor, but is not reported here to avoid unnecessary
repetition.

\subsection{Dynamic Properties}
\label{s:dynamics}

\begin{figure}[t]
  \centering
  \includegraphics[width=8.3cm]{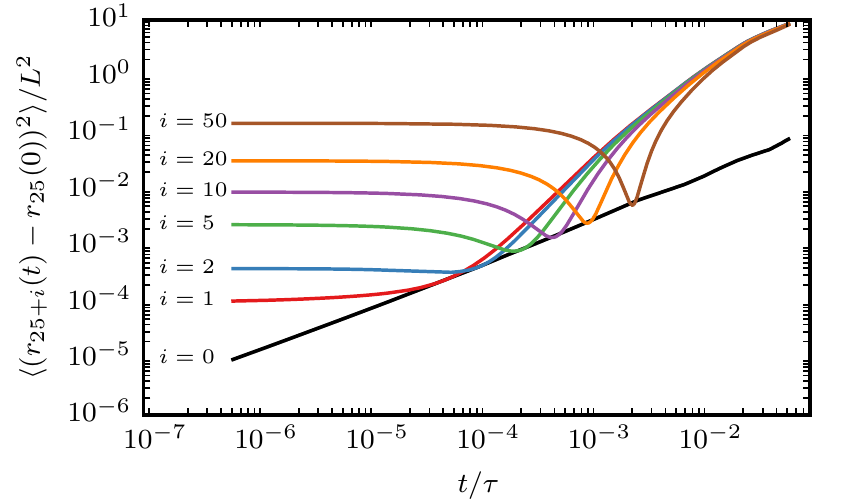}
  \caption{Mean square displacement of bead $25+i$ with respect
    to bead $j=25$ ($N=100$). $\xi_P/L=0.3$ for all lines. Black line: $Pe = 0$,
    coloured lines: $Pe = 1000$ (weak spiral regime). Black line describes
    purely diffusive motion of the leading bead $j$.
    }
  \label{f:rail}
\end{figure}

\begin{figure}[t]
\subfloat[][]{\includegraphics[width=8.3cm]{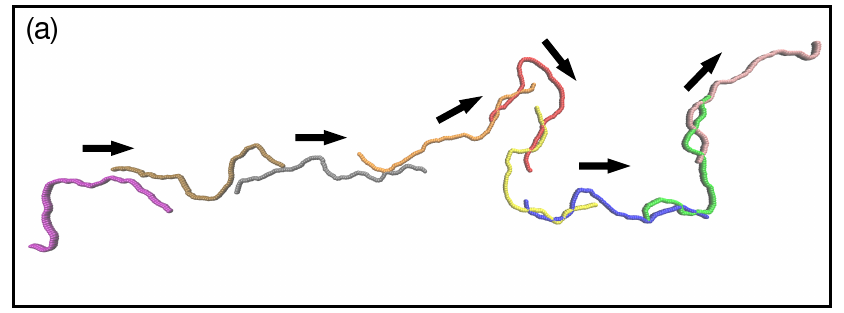}}

\subfloat[][]{\includegraphics[width=8.3cm]{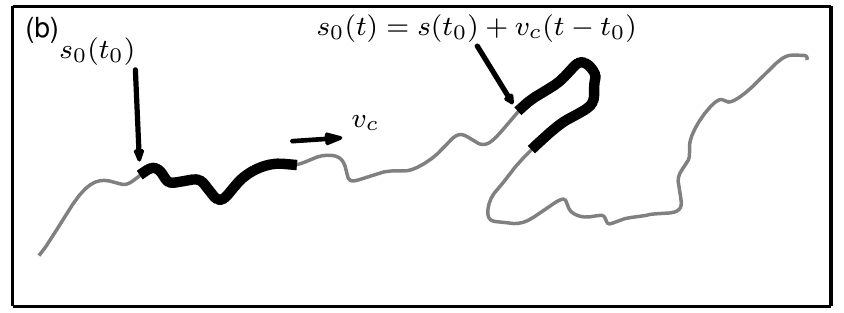}}

  \caption{(a) Series of snapshots of a filament with $\xi_P/L=0.3$ \REV{(with $N=100$)} and
    $Pe = 1000$. The chain moves along its contour superimposed with thermally activated motion.
    Colouring is to improve the distinctness of the chains.
    (b) Idealized railway motion in the absence of diffusion.
    The filament (thick black line) moves with velocity $v_c$ along the contour of an infinite chain with same $\xi_P$ (gray line). $s_0(t)$ runs along the contour of the infinite chain and marks the end point of the filament.}
  \label{f:railway_diff}
\end{figure}

The characteristic filament motion can be understood from
the mean square displacement (MSD) of a bead
$j+i$ relative to bead $j$, as shown in
Fig.~\ref{f:rail}. \REV{For comparison, the MSD of the reference
bead $j$ of a passive filament is also shown. Note that the displacement
of this bead is subdiffusive at the short lag times shown here.\cite{Grest.1986}}
Displacement functions of the
propelled beads are always larger than in the
passive case. The curves show three distinct
regimes. At small lag times,
the MSDs of active filaments display plateaus due to the
average distance of the two beads $j+i$ and $j$ along the filament.
At large lag times, the increased motion caused
by activity effects the MSDs
of the propelled beads to grow more rapidly than that
of the passive bead.
The relevant part of the MSD that shows that the characteristic filament
motion is movement along its contour is at intermediate lag times,
where the MSDs of the propelled beads pass through
minima that touch the reference MSD for \REV{thermal}
motion. At that lag time, the bead
$j+i$ has moved approximately to the position of bead $j$ at zero
lag time. The beads have moved along the chain contour, similar to the
movement of a train on a railway.
The deviation from the exact starting position of bead $j$
exactly matches the \REV{thermal} motion, which results in the
MSDs of the propelled beads touching the MSD of
the passive filament.
Thus, the characteristic movement of the filament is
motion along its contour superimposed with thermal noise, 
as depicted in Fig.~\ref{f:railway_diff}a.
Note that $\xi_P/L=0.3$ was selected in Fig.~\ref{f:rail} because
this corresponds to a rather flexible filament, where
stronger deviations from the characteristic railway motion might be expected.
This type of motion was observed in all simulations in the
polymer regime and weak spiral regime. In the
strong spiral regime, the MSDs of the propelled beads even fall below
the reference line for purely \REV{thermal} motion.

\begin{figure}[t]
\subfloat[][]{\includegraphics[width=8.3cm]{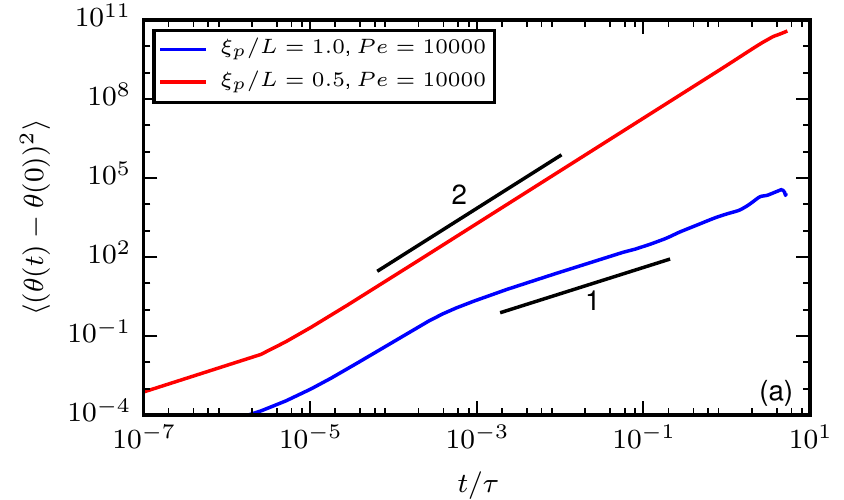}}

\subfloat[][]{\includegraphics[width=8.3cm]{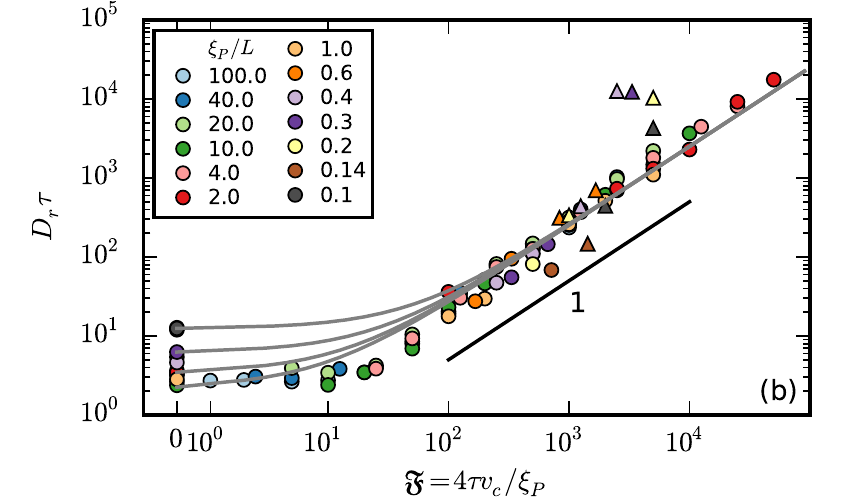}}

  \caption{(a) Mean square rotation of the end-to-end vector;
    the red line is from the strong spiral regime, the blue from
    the polymer regime ($N = 100$).
    (b) Rotational diffusion coefficient $D_r$ as a function of the flexure
    number
    $\mathfrak{F}$. Symbol shape indicates the region in the phase diagram: circle: polymer regime; triangle: weak spiral regime. Gray lines are prediction with Eq.~(\ref{e:Dr}) for different $D_{r,p}$; for the lowermost line $D_{r,p} = 9/4\,\tau^{-1}$. $N$ varies from 25 to 200 from large to small $\xi_P/L$.}
  \label{f:diff_rot}
\end{figure}

The rotational diffusion can be accessed from the orientation
of the end-to-end vector $\theta$. Its mean square rotation (MSR) is given in
Fig.~\ref{f:diff_rot}a. 
Note that complete rotations around the axis are accounted for in
our computations. $\theta(t)$ can therefore be much
larger than $2\pi$.
In both the 
spiral and the elongated state, there is a regime at short lag times
in which the MSR is dominated by the internal filament flexibility.
For the spiral state, this regime is followed by a
ballistic regime with MSR\,$\propto t^2$. For simulations in which
the spirals break up spontaneously, a subsequent regime at high
times with MSR\,$\propto t$ is expected but could not
be detected in our simulations because of finite simulation
time and strong noise at large lag times in the MSR.
For the elongated state, the regime dominated by internal flexibility is
followed by a diffusive regime with MSR\,$\propto t$. 

The rotational diffusion coefficient $D_r$ can be extracted
by fitting MSR\,$=2D_rt$ to the regime of the MSR with gradient unity on
a double--log scale.
Measured $D_r$ are given in Fig.~\ref{f:diff_rot}b
as a function of the flexure number $\mathfrak{F}$.
The diffusion coefficients collapse to
a single curve, which has a plateau at low $\mathfrak{F}$ and then grows linearly.
Strong deviations from this trend are only observed
at high $\mathfrak{F}$ when the filament is in the weak spiral regime,
and at $\mathfrak{F} = Pe = 0$ for flexible filaments,
where strong deviations from a rod-like shape increase $D_r$.

The rotational diffusion coefficient $D_r$ can be predicted
from the characteristic railway motion in
Fig.~\ref{f:railway_diff} and the relation of the
rotational diffusion coefficient to the autocorrelation
function of the end-to-end tangent vector $\mathbf{t}_e$
\begin{equation}
  \left\langle \mathbf{t}_e(t) \cdot \mathbf{t}_e(0) \right\rangle = e^{-D_rt},
  \label{e:tangent_corr_diff_rot}
\end{equation}
which is valid for lag times $t$ that are sufficiently
large such that variation of 
$\mathbf{t}_e$ is not dominated by non-diffusive behaviour at early lag times
caused by the filament flexibility
(cf. Fig.~\ref{f:diff_rot}a). With
\begin{equation}
  \mathbf{t}_e(t) = \frac{1}{L} \int_0^L\mathbf{t}(s,t)ds,
\end{equation}
where $\mathbf{t}(s,t)$ is the tangent vector at position $s$
of the filament at time $t$, the left hand side
of Eq.~(\ref{e:tangent_corr_diff_rot}) becomes
\begin{equation}
  \left\langle \mathbf{t}_e(t) \cdot \mathbf{t}_e(0)  \right\rangle = \frac{1}{L^2} 
  \int_0^L ds^\prime \int_0^L ds \left\langle \mathbf{t}(s,t) \cdot  \mathbf{t}(s\prime, 0)\right\rangle,
  \label{e:tang_integ}
\end{equation}
where the order of summations has been changed to arrive at the right-hand side
of Eq.~(\ref{e:tang_integ}).
As a representation of the characteristic railway motion (cf. Fig.~\ref{f:railway_diff}b), we write
\begin{equation}
  \mathbf{t}(s,t) = \mathbf{t}(s+v_c t, 0).
  \label{e:contour_movement}
\end{equation}
Note that this equation disregards the passive equilibrium rotation $D_{r,p}$.
With Eq.~(\ref{e:contour_movement})
and the expression for the tangent-tangent correlation
function of worm-like polymers,\cite{Kratky.1949, Saito.1967} the integrand in
Eq.~(\ref{e:tang_integ}) becomes
\begin{eqnarray}
  \left\langle \mathbf{t}(s,t) \cdot \mathbf{t}(s^\prime,0) \right\rangle &=& 
\left\langle \mathbf{t}(s+v_ct,0) \cdot \mathbf{t}(s^\prime,0)\right\rangle  \nonumber \\
     &=& \exp[-(s+v_ct - s^\prime)/\xi_P].
\end{eqnarray}
Integrating Eq.~(\ref{e:tang_integ}) provides
\begin{equation}
  \left\langle  \mathbf{t}_e(t) \cdot \mathbf{t}_e(0) \right\rangle = 
-\xi_P^2/L^2\left( e^{-v_ct/\xi_P} \left(2 - e^{L/\xi_P} - e^{-L/\xi_P} \right) \right).
\end{equation}
A second order Taylor expansion in (small) $L/\xi_P$ then gives
\begin{equation}
  \left\langle \mathbf{t}_e(t) \cdot \mathbf{t}_e(0) \right\rangle =  \exp[-v_ct/\xi_P],
\end{equation}
so that a comparison with Eq.~(\ref{e:tangent_corr_diff_rot}) finally yields
the activity-induced rotational diffusion
\begin{equation}
  D_{r,a} = v_c/\xi_P.
\end{equation}
Note that $v_c/\xi_P = \mathfrak{F}/4\tau$.
Assuming that uncorrelated activity-induced and thermal
rotation $D_{r,a}$ and $ D_{r,p}$ contribute to the overall
rotation, we write 
\begin{equation}
  D_r = D_{r,p} + D_{r,a},
  \label{e:Dr}
\end{equation}
where $D_{r,p}$ depends on $\xi_P/L$ and has the lower bound
$D_{r,p} = (9/4)\,\tau^{-1}$ for rod-like filaments.\cite{Teraoka.2002}
As can be seen from Fig.~\ref{f:diff_rot}b, Eq.~(\ref{e:Dr})
matches the simulated rotational diffusion coefficient accurately.


\begin{figure}[t]

\subfloat[][]{\includegraphics[width=8.3cm]{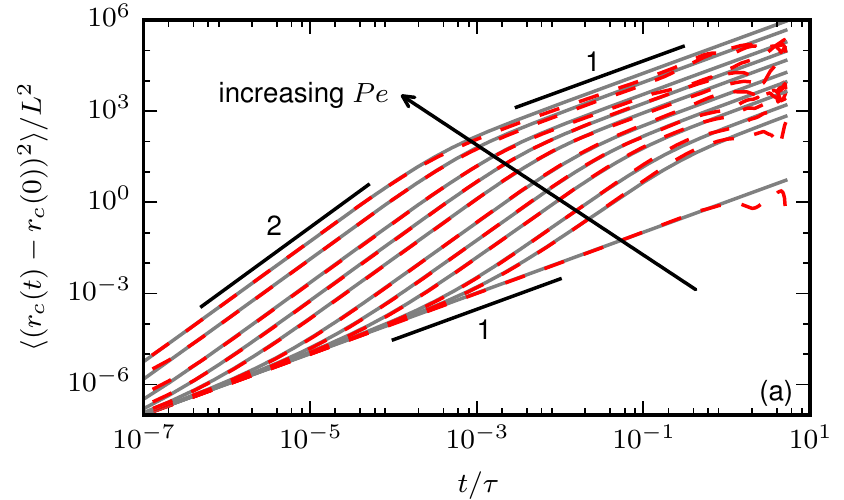}}

\subfloat[][]{\includegraphics[width=8.3cm]{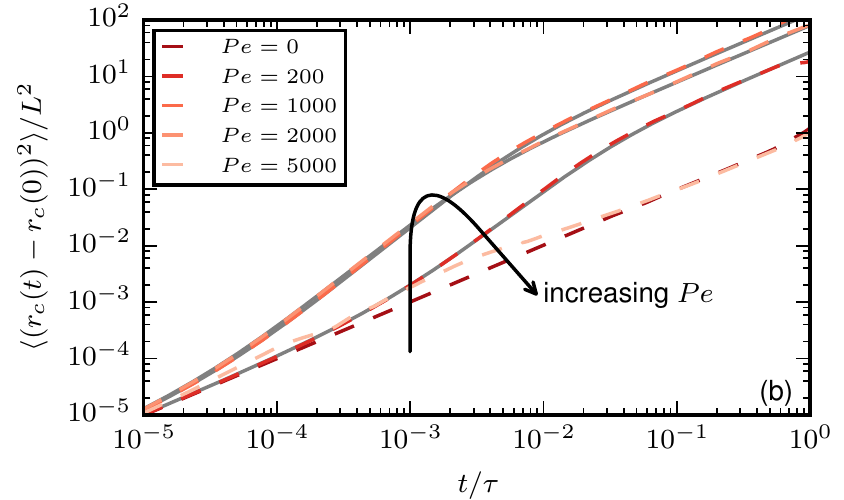}}

  \caption{Mean square displacement of the center of mass. Dashed coloured lines:
    simulation results. Gray solid lines: predictions using Eq.~(\ref{e:MSD}).
    (a) Results for $\xi_P=4.0$ (polymer regime, $N=50$) and different values of $Pe$ increasing from
    0 to 100\,000. $v_0$ is predicted from Eq.~(\ref{e:v0}), $D_r$ from Eq.~(\ref{e:Dr}) with $D_{r,p} = 9/4\,\tau^{-1}$.
    (b) Results for $\xi_P/L = 0.3$ (spirals at large $Pe$, $N=100$). $v_0$ from Eq.~(\ref{e:v0}),
    $D_r$ is determined from a fit to the measured MSD, because Eq.~(\ref{e:Dr})
    is only valid in the polymer regime.
    Spiral formation leads to a decreased MSD.}
  \label{f:MSD}
\end{figure}

The characteristics of the center-of-mass MSD
are shown in Fig.~\ref{f:MSD}. For the polymer regime, the typical
S-shape of subsequent short-time diffusive, intermediate-time ballistic,
and long-time effective diffusive behaviour develops;\cite{Zheng.2013}
stronger propulsion increases the MSD. An important difference
compared to rigid bodies is that the transition time $\tau_r = 1/D_r$
to long-time diffusive behaviour is dependent on the propulsion strength.

When spiral formation becomes important, the general trend of the MSD changes,
as shown in Fig.~\ref{f:MSD}b for a flexible
filament. In the polymer regime or weak spiral regime, increasing
$Pe$ leads to a larger displacement. In the strong spiral regime,
however, the MSD decreases. For very stable spirals, the MSD
is only weakly affected by the propulsion and almost matches
the case of purely diffusive motion.

The MSD for active point particles, spheres, or stiff rods is given by\cite{Elgeti.2015}
\begin{eqnarray}
  \langle (r_c(t) - r_c(0))^2 \rangle &=& 4D_tt + \nonumber \\
       & &     (2v_0^2/D_r^2) [D_rt + \exp (-D_rt) - 1],
  \label{e:MSD}
\end{eqnarray}
where $D_t$ is the translational diffusion coefficient and
$v_0$ is a ballistic velocity. It turns out that
Eq.~(\ref{e:MSD}) can be used to describe the MSD
for active filaments, when the three coefficients 
$D_t$, $v_0$, and $D_r$ are chosen properly.
The translational diffusion
coefficient is $D_t = L^2/4\tau = k_BT/\gamma_l L$.
We predict the rotational diffusion coefficient $D_r$ with Eq.~(\ref{e:Dr}).
Finally the effective velocity can be expressed via 
\begin{equation}
  v_0 = \frac{|\mathbf{F}_p|}{\gamma_l L}
\end{equation}
as a balance of the net external force $|\mathbf{F}_p|$ with the total friction
force $\gamma_l L v_0$. 
$|\mathbf{F}_p|$ can conveniently be expressed
as the propulsive
force per bond $f_p$ times the end-to-end vector, thus leading to
\begin{equation}
  v_0 = \frac{f_p \sqrt{\langle r_e^2 \rangle}}{\gamma_l L}.
  \label{e:v0}
\end{equation}
As shown in Fig.~\ref{f:MSD}, using these correlations for
the coefficients provides an accurate prediction of the MSD.

\begin{figure}[t]

\subfloat[][]{\includegraphics[width=8.3cm]{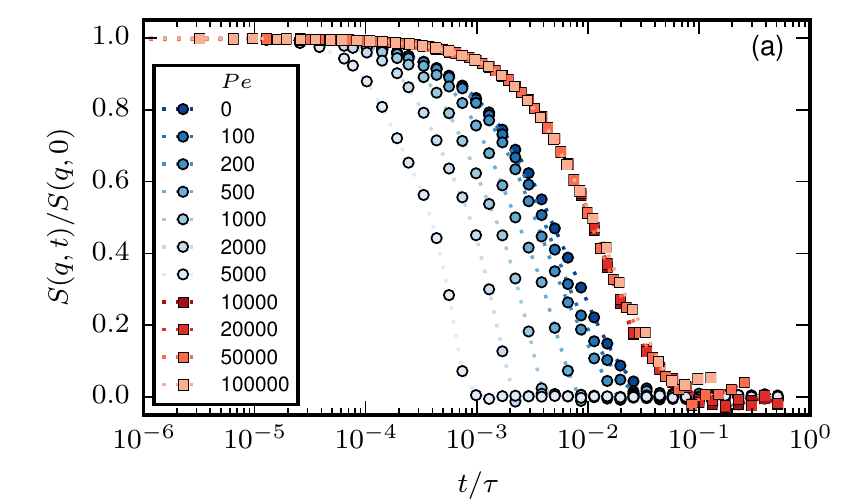}}

\subfloat[][]{\includegraphics[width=8.3cm]{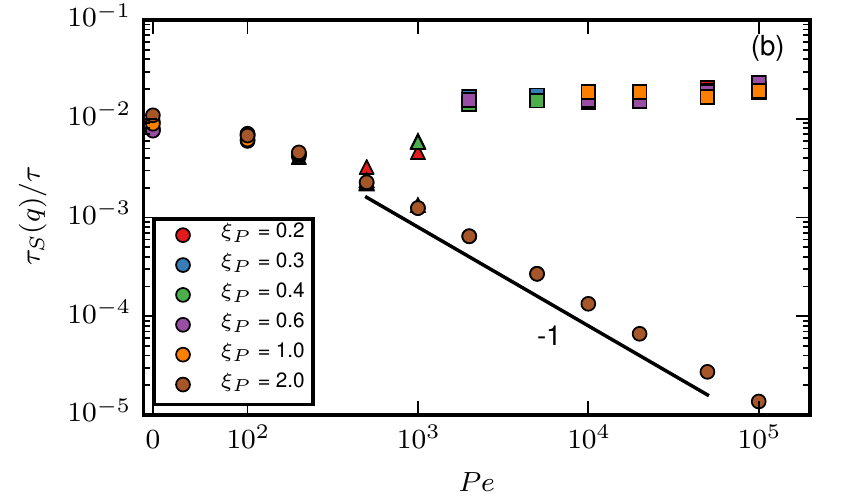}}

  \caption{(a)$S(q,t)/S(q,0)$ for $\xi_P/L = 1.0$, $N=100$, $q \approx
    5\pi/L$ and different $Pe$.
    \REV{(b) $\tau_S(q)$ for $q \approx
    5\pi/L$ .
    Circles correspond to the polymer regime, triangles to the weak
    spiral regime, and squares to
    the strong spiral regime in both plots.}
  }
  \label{f:SQT}
\end{figure}

The last item we address is the effect of propulsion on conformational
sampling. Figure~\ref{f:SQT}a shows results for the dynamic structure
factor
\begin{equation}
  S(q,t) = \left\langle  \frac{1}{N+1}\sum_{i=1}^{N+1}\sum_{j=1}^{N+1}\exp\{ i\mathbf{q} \cdot [\mathbf{r}_i(0)-\mathbf{r}_j(t)] \} \right\rangle
\end{equation}
averaged over different directions of $\mathbf{q}$. In the phase without spirals,
 $S(q,t)/S(q,0)$ decays more rapidly with increasing $Pe$, indicating a faster
change of conformations with increasing propulsion.
\REV{When spirals form, $S(q,t)/S(q,0)$ is indepenent of $Pe$ and larger
than $S(q,t)/S(q,0)$ at $Pe=0$, indicating a slow change of
conformations, which agrees with the observation of hardly any internal motion of
the chain in this regime in our simulation output.}
Note that for the strong spiral regime, the data is from simulations
where spirals formed spontaneously and did not break up.
The depicted data is a result of averaging over the spiral states
only.

\REV{To better quantify the behaviour of $S(q,t)$, we compute the
characteristic decay time of the dynamic structure factor
\begin{equation}
\tau_S(q) = \frac{\int_{t} tS(q,t) dt}{\int_{t} S(q,t) dt}.
\label{e:ts}
\end{equation}
Results for $\tau_S(q)$ at $q\approx 5\pi/L$, a $q$-vector large enough
to capture the behaviour of mainly the internal degrees of freedom,
are given in Fig~\ref{f:SQT}b.
$\tau_S(q)$ decays slowly at low $Pe$. At high $Pe$, $\tau_S$
decays inversely proportinal to $Pe$ when no spirals form
This is consistent with the picture that instantaneous conformations
are essentially identical to those of passive filaments, but they are
traversed with velocity $v_c$, corresponding to $\tau \propto Pe^{-1}$.
In the strong spiral regime, $\tau_S$ is large and independent of
$Pe$, which is a sign 
for that conformational changes are irrelevant and that $\tau_S$ is
determined by the quasi-diffusive center of mass movement.
Note that the measured $\tau_S$ at different $\xi_P/L$
collapse to a single line for both the polymer and the strong spiral regime.}

\section{Discussion}
\label{s:discussion}

The spontaneous formation of spirals is the
feature dominating the overall behaviour of self-propelled filaments, both for
dynamic and structural properties. Formation of spirals was
previously observed
for long, slender bacteria surrounded by short bacteria.\cite{Lin.2014}
It was concluded that interaction with other active
particles is a prerequisite for spiral formation. In contrast, the study
at hand shows that spirals can form even for isolated filaments,
as long as (i) the filament is sufficiently flexible, (ii)
the propulsion is sufficiently strong, and (iii) excluded
volume interactions force the tip of the filament
to wind up.

The first two conditions will be met automatically
for any real system by choosing $L$ sufficiently large
and leaving all other parameters constant 
(leading to increased $Pe$ and decreased $\xi_P/L$, i.e.,
favouring spirals).
Meeting the third condition can in general not be achieved
so easily.
A free-swimming filament in three dimensions or a
filament in two dimensions with low resistance of
crossing its own body will not form spirals. This is
also one reason why spiral formation has not yet
been observed in more experimental studies.
Agents that are similar to our model are actin filaments
or microtubules on a motility assay. The former have a
high crossing probability,\cite{Schaller.2010} 
formation of spirals is therefore not expected. The
area enclosing the actin-filament parameter space
in Fig.~\ref{f:coil_phases} must thus
be understood as that the regime where the flexibilities and
propulsion strengths permit spiral formation can 
in principle be reached
in real systems, and not so much as that sufficiently long
actin filaments will form spirals. Microtubules on dynein
carpets, which have a much lower crossing probability,\cite{Sumino.2012, Abkenar.2013}
will possibly form spirals if they are grown to
sufficient size.

Overall, except for slender bacteria,\cite{Lin.2014} we are unaware
of a microscopic example in which spiral formation was observed.
Yet, the formation of spirals is a feature that deserves more
attention in the future. First, formation of spirals is an extremely
simple non-equilibrium phenomenon that, in contrast to many
other phenomena of active matter, arises for a single self-propelled particle
and cannot easily be mapped qualitatively to passive
systems in which activity is replaced by attractive forces. It can
thus be used as a model phenomenon for the study of non-equilibrium
thermodynamics. Second, our model is very simple; a realization
in experiment seems possible within the near future. 
Finally, the formation of spirals leads
to a sudden, strong change in structural and dynamic properties.
The effect can thus potentially be used as a switch on the
microscopic scale.

\section{Summary and Outlook}
\label{s:conclusions}

We report an extensive study for the behaviour
of dilute, self-propelled, worm-like filaments in two dimensions.
The spontaneous formation and break-up of spirals
is the feature that dominates the filament
behaviour.
Spiral formation is favoured by strong propulsion and
low bending rigidity.
Propulsion has a noticeable impact
on structural properties only when spirals are dominating.
The Kratky-Porod
model\cite{Kratky.1949} is therefore valid for filaments that are
weakly propelled or have high bending rigidity.
When spiral formation becomes significant, structural properties
change drastically.

The characteristic filament motion is what we
call the railway behaviour. The chain moves along its own
contour superimposed with noise.
With the understanding of the structural properties and
the characteristic motion, 
rotational diffusion and
the center-of-mass mean square displacement can be predicted to high accuracy
when no spirals form.
In contrast to rigid bodies, propulsion has an impact on the rotational
diffusion coefficient.
Finally, propulsion enhances conformational sampling in the regime
without spirals.

An obvious next step is understanding the collective motion of such
active filaments. 
We expect that our single filament results will help to understand the
collective behaviour, which is nonetheless strongly influenced by the
additional interactions. In particular collision with other
constituents might enhance spiral formation and lead to swirl-like patterns.

\section*{Acknowledgments}
We thank Thorsten Auth for helpful discussion.
Financial support by the Deutsche Forschungsgemeinschaft
via SPP 1726 ``Microswimmers'' is gratefully
acknowledged.

\bibliography{bib.bib}

\providecommand*{\mcitethebibliography}{\thebibliography}
\csname @ifundefined\endcsname{endmcitethebibliography}
{\let\endmcitethebibliography\endthebibliography}{}
\begin{mcitethebibliography}{40}
\providecommand*{\natexlab}[1]{#1}
\providecommand*{\mciteSetBstSublistMode}[1]{}
\providecommand*{\mciteSetBstMaxWidthForm}[2]{}
\providecommand*{\mciteBstWouldAddEndPuncttrue}
  {\def\EndOfBibitem{\unskip.}}
\providecommand*{\mciteBstWouldAddEndPunctfalse}
  {\let\EndOfBibitem\relax}
\providecommand*{\mciteSetBstMidEndSepPunct}[3]{}
\providecommand*{\mciteSetBstSublistLabelBeginEnd}[3]{}
\providecommand*{\EndOfBibitem}{}
\mciteSetBstSublistMode{f}
\mciteSetBstMaxWidthForm{subitem}
{(\emph{\alph{mcitesubitemcount}})}
\mciteSetBstSublistLabelBeginEnd{\mcitemaxwidthsubitemform\space}
{\relax}{\relax}

\bibitem[Elgeti \emph{et~al.}(2015)Elgeti, Winkler, and Gompper]{Elgeti.2015}
J.~Elgeti, R.~G. Winkler and G.~Gompper, \emph{Rep. Prog. Phys.}, 2015,
  \textbf{78}, 056601\relax
\mciteBstWouldAddEndPuncttrue
\mciteSetBstMidEndSepPunct{\mcitedefaultmidpunct}
{\mcitedefaultendpunct}{\mcitedefaultseppunct}\relax
\EndOfBibitem
\bibitem[Marchetti \emph{et~al.}(2013)Marchetti, Joanny, Ramaswamy, Liverpool,
  Prost, Rao, and Simha]{Marchetti.2013}
M.~C. Marchetti, J.~F. Joanny, S.~Ramaswamy, T.~B. Liverpool, J.~Prost, M.~Rao
  and R.~A. Simha, \emph{Rev. Mod. Phys.}, 2013, \textbf{85}, 1143--1189\relax
\mciteBstWouldAddEndPuncttrue
\mciteSetBstMidEndSepPunct{\mcitedefaultmidpunct}
{\mcitedefaultendpunct}{\mcitedefaultseppunct}\relax
\EndOfBibitem
\bibitem[Cates and MacKintosh(2011)]{Cates.2011}
M.~E. Cates and F.~C. MacKintosh, \emph{Soft Matter}, 2011, \textbf{7},
  3050--3051\relax
\mciteBstWouldAddEndPuncttrue
\mciteSetBstMidEndSepPunct{\mcitedefaultmidpunct}
{\mcitedefaultendpunct}{\mcitedefaultseppunct}\relax
\EndOfBibitem
\bibitem[Rodriguez \emph{et~al.}(2003)Rodriguez, Schaefer, Mandato, Forscher,
  Bement, and Waterman-Storer]{Rodriguez.2003}
O.~C. Rodriguez, A.~W. Schaefer, C.~A. Mandato, P.~Forscher, W.~M. Bement and
  C.~M. Waterman-Storer, \emph{Nat. Cell Biol.}, 2003, \textbf{5},
  599--609\relax
\mciteBstWouldAddEndPuncttrue
\mciteSetBstMidEndSepPunct{\mcitedefaultmidpunct}
{\mcitedefaultendpunct}{\mcitedefaultseppunct}\relax
\EndOfBibitem
\bibitem[Rashedul~Kabir \emph{et~al.}(2012)Rashedul~Kabir, Wada, Inoue, Tamura,
  Kajihara, Mayama, Sada, Kakugo, and Gong]{RashedulKabir.2012}
A.~M. Rashedul~Kabir, S.~Wada, D.~Inoue, Y.~Tamura, T.~Kajihara, H.~Mayama,
  K.~Sada, A.~Kakugo and J.~P. Gong, \emph{Soft Matter}, 2012, \textbf{8},
  10863--10867\relax
\mciteBstWouldAddEndPuncttrue
\mciteSetBstMidEndSepPunct{\mcitedefaultmidpunct}
{\mcitedefaultendpunct}{\mcitedefaultseppunct}\relax
\EndOfBibitem
\bibitem[Sumino \emph{et~al.}(2012)Sumino, Nagai, Shitaka, Tanaka, Yoshikawa,
  Chat\'{e}, and Oiwa]{Sumino.2012}
Y.~Sumino, K.~H. Nagai, Y.~Shitaka, D.~Tanaka, K.~Yoshikawa, H.~Chat\'{e} and
  K.~Oiwa, \emph{Nature}, 2012, \textbf{483}, 448--452\relax
\mciteBstWouldAddEndPuncttrue
\mciteSetBstMidEndSepPunct{\mcitedefaultmidpunct}
{\mcitedefaultendpunct}{\mcitedefaultseppunct}\relax
\EndOfBibitem
\bibitem[Lin \emph{et~al.}(2014)Lin, Lo, and Lo]{Lin.2014}
S.-N. Lin, W.-C. Lo and C.-J. Lo, \emph{Soft Matter}, 2014, \textbf{10},
  760--766\relax
\mciteBstWouldAddEndPuncttrue
\mciteSetBstMidEndSepPunct{\mcitedefaultmidpunct}
{\mcitedefaultendpunct}{\mcitedefaultseppunct}\relax
\EndOfBibitem
\bibitem[Sasaki \emph{et~al.}(2014)Sasaki, Takikawa, Jampani, Hoshikawa, Seto,
  Bahr, Herminghaus, Hidaka, and Orihara]{Sasaki.2014}
Y.~Sasaki, Y.~Takikawa, V.~S.~R. Jampani, H.~Hoshikawa, T.~Seto, C.~Bahr,
  S.~Herminghaus, Y.~Hidaka and H.~Orihara, \emph{Soft Matter}, 2014,
  \textbf{10}, 8813--8820\relax
\mciteBstWouldAddEndPuncttrue
\mciteSetBstMidEndSepPunct{\mcitedefaultmidpunct}
{\mcitedefaultendpunct}{\mcitedefaultseppunct}\relax
\EndOfBibitem
\bibitem[Vach and Faivre(2015)]{Vach.2015}
P.~J. Vach and D.~Faivre, \emph{Sci. Rep.}, 2015, \textbf{5}, 9364\relax
\mciteBstWouldAddEndPuncttrue
\mciteSetBstMidEndSepPunct{\mcitedefaultmidpunct}
{\mcitedefaultendpunct}{\mcitedefaultseppunct}\relax
\EndOfBibitem
\bibitem[Sanchez \emph{et~al.}(2011)Sanchez, Welch, Nicastro, and
  Dogic]{Sanchez.2011}
T.~Sanchez, D.~Welch, D.~Nicastro and Z.~Dogic, \emph{Science}, 2011,
  \textbf{333}, 456--459\relax
\mciteBstWouldAddEndPuncttrue
\mciteSetBstMidEndSepPunct{\mcitedefaultmidpunct}
{\mcitedefaultendpunct}{\mcitedefaultseppunct}\relax
\EndOfBibitem
\bibitem[Sanchez \emph{et~al.}(2012)Sanchez, Chen, DeCamp, Heymann, and
  Dogic]{Sanchez.2012}
T.~Sanchez, D.~T.~N. Chen, S.~J. DeCamp, M.~Heymann and Z.~Dogic,
  \emph{Nature}, 2012, \textbf{491}, 431--434\relax
\mciteBstWouldAddEndPuncttrue
\mciteSetBstMidEndSepPunct{\mcitedefaultmidpunct}
{\mcitedefaultendpunct}{\mcitedefaultseppunct}\relax
\EndOfBibitem
\bibitem[Sekimoto \emph{et~al.}(1995)Sekimoto, Mori, Tawada, and
  Toyoshima]{Sekimoto.1995}
K.~Sekimoto, N.~Mori, K.~Tawada and Y.~Toyoshima, \emph{Phys. Rev. Lett.},
  1995, \textbf{75}, 172--175\relax
\mciteBstWouldAddEndPuncttrue
\mciteSetBstMidEndSepPunct{\mcitedefaultmidpunct}
{\mcitedefaultendpunct}{\mcitedefaultseppunct}\relax
\EndOfBibitem
\bibitem[Bourdieu \emph{et~al.}(1995)Bourdieu, Duke, Elowitz, Winkelmann,
  Leibler, and Libchaber]{Bourdieu.1995}
L.~Bourdieu, T.~Duke, M.~Elowitz, D.~Winkelmann, S.~Leibler and A.~Libchaber,
  \emph{Phys. Rev. Lett.}, 1995, \textbf{75}, 176--179\relax
\mciteBstWouldAddEndPuncttrue
\mciteSetBstMidEndSepPunct{\mcitedefaultmidpunct}
{\mcitedefaultendpunct}{\mcitedefaultseppunct}\relax
\EndOfBibitem
\bibitem[Laskar \emph{et~al.}(2013)Laskar, Singh, Ghose, Jayaraman, Kumar, and
  Adhikari]{Laskar.2013}
A.~Laskar, R.~Singh, S.~Ghose, G.~Jayaraman, P.~B.~S. Kumar and R.~Adhikari,
  \emph{Sci. Rep.}, 2013, \textbf{3}, 1964\relax
\mciteBstWouldAddEndPuncttrue
\mciteSetBstMidEndSepPunct{\mcitedefaultmidpunct}
{\mcitedefaultendpunct}{\mcitedefaultseppunct}\relax
\EndOfBibitem
\bibitem[Chelakkot \emph{et~al.}(2014)Chelakkot, Gopinath, Mahadevan, and
  Hagan]{Chelakkot.2014}
R.~Chelakkot, A.~Gopinath, L.~Mahadevan and M.~F. Hagan, \emph{J. R. Soc.
  Interface}, 2014, \textbf{11}, 20130884\relax
\mciteBstWouldAddEndPuncttrue
\mciteSetBstMidEndSepPunct{\mcitedefaultmidpunct}
{\mcitedefaultendpunct}{\mcitedefaultseppunct}\relax
\EndOfBibitem
\bibitem[Farkas \emph{et~al.}(2002)Farkas, Der\'enyi, and Vicsek]{Farkas.2002}
Z.~Farkas, I.~Der\'enyi and T.~Vicsek, in \emph{Structure and Dynamics of
  Confined Polymers}, ed. J.~J. Kasianowicz, M.~S.~Z. Kellermayer and D.~W.
  Deamer, Springer Netherlands, 2002, vol.~87, pp. 327--332\relax
\mciteBstWouldAddEndPuncttrue
\mciteSetBstMidEndSepPunct{\mcitedefaultmidpunct}
{\mcitedefaultendpunct}{\mcitedefaultseppunct}\relax
\EndOfBibitem
\bibitem[Jayaraman \emph{et~al.}(2012)Jayaraman, Ramachandran, Ghose, Laskar,
  Bhamla, Kumar, and Adhikari]{Jayaraman.2012}
G.~Jayaraman, S.~Ramachandran, S.~Ghose, A.~Laskar, M.~S. Bhamla, P.~B.~S.
  Kumar and R.~Adhikari, \emph{Phys. Rev. Lett.}, 2012, \textbf{109},
  158302\relax
\mciteBstWouldAddEndPuncttrue
\mciteSetBstMidEndSepPunct{\mcitedefaultmidpunct}
{\mcitedefaultendpunct}{\mcitedefaultseppunct}\relax
\EndOfBibitem
\bibitem[Jiang and Hou(2014)]{Jiang.2014_1}
H.~Jiang and Z.~Hou, \emph{Soft Matter}, 2014, \textbf{10}, 9248--53\relax
\mciteBstWouldAddEndPuncttrue
\mciteSetBstMidEndSepPunct{\mcitedefaultmidpunct}
{\mcitedefaultendpunct}{\mcitedefaultseppunct}\relax
\EndOfBibitem
\bibitem[Jiang and Hou(2014)]{Jiang.2014_2}
H.~Jiang and Z.~Hou, \emph{Soft Matter}, 2014, \textbf{10}, 1012--1017\relax
\mciteBstWouldAddEndPuncttrue
\mciteSetBstMidEndSepPunct{\mcitedefaultmidpunct}
{\mcitedefaultendpunct}{\mcitedefaultseppunct}\relax
\EndOfBibitem
\bibitem[Ghosh and Gov(2014)]{Ghosh.2014}
A.~Ghosh and N.~S. Gov, \emph{Biophys. J.}, 2014, \textbf{107},
  1065--1073\relax
\mciteBstWouldAddEndPuncttrue
\mciteSetBstMidEndSepPunct{\mcitedefaultmidpunct}
{\mcitedefaultendpunct}{\mcitedefaultseppunct}\relax
\EndOfBibitem
\bibitem[Kaiser \emph{et~al.}(2015)Kaiser, Babel, ten Hagen, von Ferber, and
  L\"{o}wen]{Kaiser.2015}
A.~Kaiser, S.~Babel, B.~ten Hagen, C.~von Ferber and H.~L\"{o}wen, \emph{J.
  Chem. Phys.}, 2015, \textbf{142}, 124905\relax
\mciteBstWouldAddEndPuncttrue
\mciteSetBstMidEndSepPunct{\mcitedefaultmidpunct}
{\mcitedefaultendpunct}{\mcitedefaultseppunct}\relax
\EndOfBibitem
\bibitem[D\"unweg and Paul(1991)]{Dunweg.1991}
B.~D\"unweg and W.~Paul, \emph{Int. J. Mod. Phys. C}, 1991, \textbf{02},
  817--827\relax
\mciteBstWouldAddEndPuncttrue
\mciteSetBstMidEndSepPunct{\mcitedefaultmidpunct}
{\mcitedefaultendpunct}{\mcitedefaultseppunct}\relax
\EndOfBibitem
\bibitem[Downton and Stark(2009)]{Downton.2009}
M.~T. Downton and H.~Stark, \emph{J. Phys. Condens. Matter}, 2009, \textbf{21},
  204101\relax
\mciteBstWouldAddEndPuncttrue
\mciteSetBstMidEndSepPunct{\mcitedefaultmidpunct}
{\mcitedefaultendpunct}{\mcitedefaultseppunct}\relax
\EndOfBibitem
\bibitem[G\"otze and Gompper(2010)]{Goetze.2010}
I.~O. G\"otze and G.~Gompper, \emph{Phys. Rev. E}, 2010, \textbf{82},
  041921\relax
\mciteBstWouldAddEndPuncttrue
\mciteSetBstMidEndSepPunct{\mcitedefaultmidpunct}
{\mcitedefaultendpunct}{\mcitedefaultseppunct}\relax
\EndOfBibitem
\bibitem[Z\"ottl and Stark(2014)]{Zoettl.2014}
A.~Z\"ottl and H.~Stark, \emph{Phys. Rev. Lett.}, 2014, \textbf{112},
  118101\relax
\mciteBstWouldAddEndPuncttrue
\mciteSetBstMidEndSepPunct{\mcitedefaultmidpunct}
{\mcitedefaultendpunct}{\mcitedefaultseppunct}\relax
\EndOfBibitem
\bibitem[Elgeti and Gompper(2009)]{Elgeti.2009}
J.~Elgeti and G.~Gompper, \emph{Europhys. Lett.}, 2009, \textbf{85},
  38002\relax
\mciteBstWouldAddEndPuncttrue
\mciteSetBstMidEndSepPunct{\mcitedefaultmidpunct}
{\mcitedefaultendpunct}{\mcitedefaultseppunct}\relax
\EndOfBibitem
\bibitem[Drescher \emph{et~al.}(2011)Drescher, Dunkel, Cisneros, Ganguly, and
  Goldstein]{Drescher.2011}
K.~Drescher, J.~Dunkel, L.~H. Cisneros, S.~Ganguly and R.~Goldstein,
  \emph{Proc. Natl. Acad. Sci. USA}, 2011, \textbf{108}, 10940--10945\relax
\mciteBstWouldAddEndPuncttrue
\mciteSetBstMidEndSepPunct{\mcitedefaultmidpunct}
{\mcitedefaultendpunct}{\mcitedefaultseppunct}\relax
\EndOfBibitem
\bibitem[Gray and Lissmann(1964)]{Gray.1964}
J.~Gray and H.~W. Lissmann, \emph{J. Exp. Biol.}, 1964, \textbf{41},
  135--154\relax
\mciteBstWouldAddEndPuncttrue
\mciteSetBstMidEndSepPunct{\mcitedefaultmidpunct}
{\mcitedefaultendpunct}{\mcitedefaultseppunct}\relax
\EndOfBibitem
\bibitem[Korta \emph{et~al.}(2007)Korta, Clark, Gabel, Mahadevan, and
  Samuel]{Korta.2007}
J.~Korta, D.~A. Clark, C.~V. Gabel, L.~Mahadevan and A.~D.~T. Samuel, \emph{J.
  Exp. Biol.}, 2007, \textbf{210}, 2383--2389\relax
\mciteBstWouldAddEndPuncttrue
\mciteSetBstMidEndSepPunct{\mcitedefaultmidpunct}
{\mcitedefaultendpunct}{\mcitedefaultseppunct}\relax
\EndOfBibitem
\bibitem[Kratky and Porod(1949)]{Kratky.1949}
O.~Kratky and G.~Porod, \emph{J. Colloid Sci.}, 1949, \textbf{4}, 35--70\relax
\mciteBstWouldAddEndPuncttrue
\mciteSetBstMidEndSepPunct{\mcitedefaultmidpunct}
{\mcitedefaultendpunct}{\mcitedefaultseppunct}\relax
\EndOfBibitem
\bibitem[Sait\^{o} \emph{et~al.}(1967)Sait\^{o}, Takahashi, and
  Yunoki]{Saito.1967}
N.~Sait\^{o}, K.~Takahashi and Y.~Yunoki, \emph{J. Phys. Soc. Jpn.}, 1967,
  \textbf{22}, 219--226\relax
\mciteBstWouldAddEndPuncttrue
\mciteSetBstMidEndSepPunct{\mcitedefaultmidpunct}
{\mcitedefaultendpunct}{\mcitedefaultseppunct}\relax
\EndOfBibitem
\bibitem[Plimpton(1995)]{Plimpton.1995}
S.~Plimpton, \emph{J. Comput. Phys.}, 1995, \textbf{117}, 1--19\relax
\mciteBstWouldAddEndPuncttrue
\mciteSetBstMidEndSepPunct{\mcitedefaultmidpunct}
{\mcitedefaultendpunct}{\mcitedefaultseppunct}\relax
\EndOfBibitem
\bibitem[Yang \emph{et~al.}(2010)Yang, Marceau, and Gompper]{Yang.2010}
Y.~Yang, V.~Marceau and G.~Gompper, \emph{Phys. Rev. E}, 2010, \textbf{82},
  031904\relax
\mciteBstWouldAddEndPuncttrue
\mciteSetBstMidEndSepPunct{\mcitedefaultmidpunct}
{\mcitedefaultendpunct}{\mcitedefaultseppunct}\relax
\EndOfBibitem
\bibitem[Abkenar \emph{et~al.}(2013)Abkenar, Marx, Auth, and
  Gompper]{Abkenar.2013}
M.~Abkenar, K.~Marx, T.~Auth and G.~Gompper, \emph{Phys. Rev. E}, 2013,
  \textbf{88}, 062314\relax
\mciteBstWouldAddEndPuncttrue
\mciteSetBstMidEndSepPunct{\mcitedefaultmidpunct}
{\mcitedefaultendpunct}{\mcitedefaultseppunct}\relax
\EndOfBibitem
\bibitem[Grest and Kremer(1986)]{Grest.1986}
G.~S. Grest and K.~Kremer, \emph{Phys. Rev. A}, 1986, \textbf{33},
  3628--3631\relax
\mciteBstWouldAddEndPuncttrue
\mciteSetBstMidEndSepPunct{\mcitedefaultmidpunct}
{\mcitedefaultendpunct}{\mcitedefaultseppunct}\relax
\EndOfBibitem
\bibitem[Teraoka(2002)]{Teraoka.2002}
I.~Teraoka, \emph{Polymer Solutions: An Introduction to Physical Properties},
  John Wiley \& Sons, Inc., New York, 2002\relax
\mciteBstWouldAddEndPuncttrue
\mciteSetBstMidEndSepPunct{\mcitedefaultmidpunct}
{\mcitedefaultendpunct}{\mcitedefaultseppunct}\relax
\EndOfBibitem
\bibitem[Zheng \emph{et~al.}(2013)Zheng, ten Hagen, Kaiser, Wu, Cui, Silber-Li,
  and L\"{o}wen]{Zheng.2013}
X.~Zheng, B.~ten Hagen, A.~Kaiser, M.~Wu, H.~Cui, Z.~Silber-Li and
  H.~L\"{o}wen, \emph{Phys. Rev. E}, 2013, \textbf{88}, 032304\relax
\mciteBstWouldAddEndPuncttrue
\mciteSetBstMidEndSepPunct{\mcitedefaultmidpunct}
{\mcitedefaultendpunct}{\mcitedefaultseppunct}\relax
\EndOfBibitem
\bibitem[Schaller \emph{et~al.}(2010)Schaller, Weber, Semmrich, Frey, and
  Bausch]{Schaller.2010}
V.~Schaller, C.~Weber, C.~Semmrich, E.~Frey and A.~R. Bausch, \emph{Nature},
  2010, \textbf{467}, 73--77\relax
\mciteBstWouldAddEndPuncttrue
\mciteSetBstMidEndSepPunct{\mcitedefaultmidpunct}
{\mcitedefaultendpunct}{\mcitedefaultseppunct}\relax
\EndOfBibitem
\end{mcitethebibliography}
\bibliographystyle{rsc} 

\end{document}